\documentclass[useAMS,usenatbib]{mn2e}
\usepackage{epsfig}
\usepackage{color}

\title[Kinematics of Andromeda's South-West Cloud]{Accretion in action: phase space coherence of stellar debris and globular clusters in Andromeda's South-West Cloud\thanks{The data presented herein were obtained at the W.M. Keck Observatory, which is operated as a scientific partnership among the California Institute of Technology, the University of California and the National Aeronautics and Space Administration. The Observatory was made possible by the generous financial support of the W.M. Keck Foundation.}}
\author[Mackey et al.]{A. D. Mackey$^1$, G. F. Lewis$^2$, M. L. M. Collins$^3$, N. F. Bate$^2$, R. A. Ibata$^4$, 
\newauthor N. F. Martin$^{3,4}$, S. Chapman$^{5,6}$, A. Conn$^2$, P. Elahi$^2$, A. M. N. Ferguson$^7$, A. Huxor$^8$, 
\newauthor M. Irwin$^6$, A. McConnachie$^9$, B. McMonigal$^2$, J. Pe\~{n}arrubia$^7$, J. Veljanoski$^7$
\\
$^1$RSAA, Australian National University, Mt. Stromlo Observatory, Cotter Road, Weston Creek, ACT 2611, Australia\\
$^2$Sydney Institute for Astronomy, School of Physics A28, University of Sydney, NSW 2006, Australia\\
$^3$Max-Planck-Institut f\"{u}r Astronomie, K\"{o}nigstuhl 17, D-69117 Heidelberg, Germany\\
$^4$Observatoire Astronomique de Strasbourg, Universit\'{e} de Strasbourg, CNRS, 11 rue de l'Universit\'{e}, F-67000 Strasbourg, France\\
$^5$Department of Physics and Atmospheric Science, Dalhousie University, 6310 Coburg Road, Halifax, Nova Scotia B3H 4R2, Canada\\
$^6$Institute of Astronomy, University of Cambridge, Madingley Road, Cambridge CB3 0HA\\
$^7$Institute for Astronomy, University of Edinburgh, Royal Observatory, Blackford Hill, Edinburgh EH9 3HJ\\
$^8$Astronomisches Rechen-Institut, Universit\"{a}t Heidelberg, M\"{o}nchhofstra{\ss}e 12 - 14, D-69120 Heidelberg, Germany\\
$^9$NRC Herzberg Institute of Astrophysics, 5071 West Saanich Road, Victoria, British Columbia V9E 2E7, Canada
}

\begin{document}


\pagerange{\pageref{firstpage}--\pageref{lastpage}} \pubyear{2014}

\maketitle

\label{firstpage}

\begin{abstract}
A central tenet of the current cosmological paradigm is that galaxies grow over time through the accretion of smaller systems. Here, we present new kinematic measurements near the centre of one of the densest pronounced substructures, the South-West Cloud, in the outer halo of our nearest giant neighbour, the Andromeda galaxy. These observations reveal that the kinematic properties of this region of the South-West Cloud are consistent with those of PA-8, a globular cluster previously shown to be co-spatial with the stellar substructure. In this sense the situation is reminiscent of the handful of globular clusters that sit near the heart of the Sagittarius dwarf galaxy, a system that is currently being accreted into the Milky Way, confirming that accretion deposits not only stars but also globular clusters into the halos of large galaxies.
\end{abstract}


\section{Introduction}
The halo of M31, as revealed in the Pan-Andromeda Archaeological Survey \citep[PAndAS;][]{mcconnachie:09}, is strewn with substructure out to a projected radius of at least $150$ kpc. PAndAS was conceived to take advantage of the proximity of M31 -- close enough to resolve individual stars, far enough away for a panoramic view -- to study galaxy assembly in a system comparable to our own Milky Way.

Much recent work on M31 halo substructure has focussed on dwarf galaxies and globular clusters (GCs). The population of known M31 dwarf satellites has grown to over thirty \citep[see e.g.,][and references therein]{mcconnachie:12,martin:13a,martin:13b,martin:13c}, sixteen of which were discovered in PAndAS data \citep[e.g.,][]{martin:06,mcconnachie:08,richardson:11}. The population of known halo globular clusters has also grown very substantially \citep{huxor:08,huxor:14}. Of particular interest here, M31 outer halo globular clusters exhibit a striking spatial coincidence with diffuse stellar streams and overdensities in the field \citep[][2014 in prep.]{mackey:10}. Together with the observation that subgroups of GCs on individual streams tend to possess correlated velocities \citep{veljanoski:14}, this implies that a significant fraction of the remote GC system of M31 has been acquired via the accretion and disruption of dwarf galaxies hosting their own small GC retinues.

Although various stellar overdensities within the inner M31 halo have been extensively probed \citep[e.g.,][]{ibata:04,ferguson:05,gilbert:07,gilbert:09,richardson:08,fardal:12}, the only discrete substructure in the remote outskirts for which the constituent stellar populations have been studied in detail is the South-West Cloud \citep[SWC,][]{bate:14}. This feature, sitting at a projected galactocentric radius $R_p \sim 90$ kpc and spanning a $\sim 50$ kpc arc on the sky, is diffuse with a mean surface brightness $\Sigma_V = 31.7\pm0.3$ mag arcsec$^{-2}$. Via the luminosity-metallicity relation \citep{kirby:11}, its mean $[$Fe$/$H$] \approx -1.3$ implies that it was amongst M31's brighter satellites ($M_V \sim -12.4$) prior to its disruption. The presence of three globular clusters along the stream \citep[PA-7, PA-8 and PA-14,][]{mackey:13,huxor:14} is also consistent with a moderately luminous progenitor. 

While the available evidence points towards the SWC being a stellar stream from a dissolving dwarf galaxy, testing this hypothesis requires velocity information. As a starting point, the kinematics of the three overlying GCs imply, under the assumption that they are members of the substructure, a velocity gradient of $\ga 70$\,km\,s$^{-1}$ along the main body of the SWC \citep{veljanoski:14}. By measuring the velocities of SWC stars near these GCs, we can obtain definitive confirmation that they are indeed associated with the substructure. Such confirmation is rare -- the only {\it unambiguous} examples known in the Milky Way or M31 are the GCs associated with the Sagittarius dwarf \citep[e.g.,][]{dacosta:95,bellazzini:03,law:10}, and the cluster HEC12\footnote{Alternatively EC4 \citep[see][]{mackey:06,huxor:08}.} which is kinematically tied to the metal-poor component of Stream C in the M31 halo \citep{chapman:08,collins:09}.  

To date, the velocity of the SWC has been determined at only one location. \citet{gilbert:12} serendipitously discovered a cold kinematic peak at $v_r=-373.5 \pm 3.0$\,km\,s$^{-1}$ whilst investigating the velocities of M31 halo stars, in a field lying near the putative SWC globular cluster PA-14. The observed velocity of that cluster, at $v_r=-363 \pm 9$\,km\,s$^{-1}$ \citep{veljanoski:14}, suggests a kinematic link. In this paper, we present a new detection of the SWC velocity and dispersion at a location near its central density peak, and close to the globular cluster PA-8. 

\section{Observations \& Data Reduction}
\label{s:data}
\begin{figure}
\centering
\includegraphics[height=76mm]{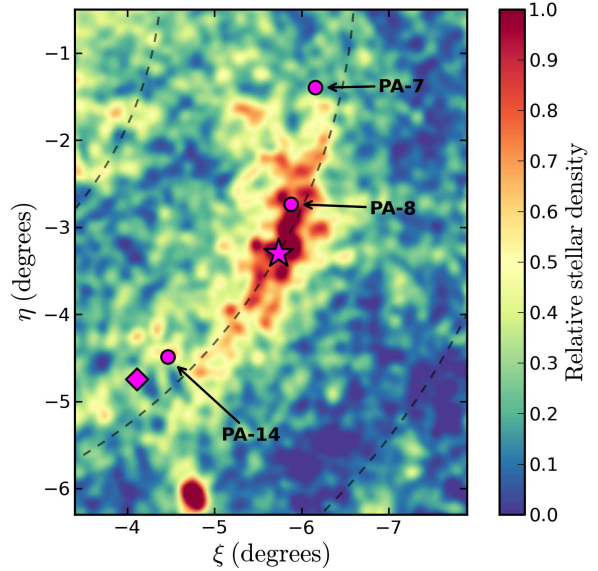}
\caption{PAndAS map of  the SWC, showing the surface density of RGB stars falling near the colour-magnitude sequence belonging to the substructure (see Figure \ref{f:cmd+vel}). Star-counts were conducted in $2\arcmin \times 2\arcmin$ bins and then convolved with a Gaussian  of $\sigma = 3\arcmin$. Contamination from foreground stars and background galaxies was subtracted using the model developed by \citet{martin:13b}. The colour-map has been scaled relative to the peak density in the central parts of the substructure. The three globular clusters projected onto the SWC  (magenta circles) are labelled. Our DEIMOS field is indicated with a star; that from \citet{gilbert:12} is marked with a diamond. The coordinates $(\xi, \eta)$ represent a tangent plane centred on M31. Projected galactocentric distances of $60$, $90$, and $120$ kpc are marked with dashed lines. The overdensity to the south is the dwarf galaxy And XIX.}\label{f:map}
\end{figure}

We used the DEep Imaging Multi-Object Spectrograph (DEIMOS) instrument on the 10m Keck II telescope to obtain spectra of candidate members of the SWC by means of a single multi-slit mask placed at $(\alpha, \delta) = (00^{\rm h}13^{\rm m}49\fs77, +37\degr44\arcmin54\farcs9)$ (J2000.0), approximately $35\arcmin$ south of PA-8. Figure \ref{f:map} shows the location of the field. 

\subsection{Target selection}
\label{ss:targ}
We selected targets from the PAndAS photometric catalogue, the details of which have previously been described by \citet{martin:13b} and \citet{ibata:14}. We isolated all stars lying within the main body of the SWC as defined by \citet[][their Figure 3]{bate:14} and inside the box on the colour-magnitude diagram (CMD) with limits $0.2 \le (g-i)_0 \le 2.8$ and $20.5 \le i_0 \le 24.0$. We then subtracted the contamination model developed by \citet{martin:13b} to remove contributions due to Milky Way foreground stars and unresolved background galaxies, revealing the sequence due to red-giant branch (RGB) stars in the SWC (see Figure \ref{f:cmd+vel}). This narrow CMD region is closely bounded by Dartmouth isochrones \citep{dotter:08} with $[$Fe$/$H$] = -1.8$ and $-1.1$; we used the area so defined to identify the most likely members of the substructure in the vicinity of its density peak.
We assigned a numerical priority to each such star based on (i) brightness, and (ii) proximity to the RGB sequence, and then optimised the position and orientation of the DEIMOS mask by maximising the sum of the priority values underneath it.  Finally, we filled out the remainder of the mask with Milky Way foreground stars.

In total we targeted $56$ stars, of which $41$ were selected as candidate RGB members of the SWC and $15$ as Milky Way foreground stars. Figure \ref{f:cmd+vel} shows the positions of our targets on the CMD. The shape of the SWC metallicity distribution function \citep[MDF,][]{bate:14} is very similar to that of the smooth halo component of M31 at commensurate galactocentric radii \citep{ibata:14}.  Our photometric selection procedure thus cannot eliminate contamination of the SWC sample by RGB members of the underlying M31 halo; this is discussed further in Section \ref{s:analysis}.

\subsection{Data acquisition}
Our data were obtained on 2013 September 11 (Program 2013B-Z297D; PI: Mackey). Atmospheric conditions were clear and stable with seeing $\approx 0.8\arcsec$; however, due to high humidity we managed a total integration of only $3540$s over three sub-exposures before the telescope was closed -- about half of what we planned. We used a mask with $1\arcsec$ slits, and employed the high-resolution $1200$ line mm$^{-1}$ grating and OG550 blocking filter, at a central wavelength of $7800$\AA. This provides spectral coverage of $\sim 6500-9200$\AA\ with a dispersion of $0.33$\AA\ pixel$^{-1}$, and a resolution of $\approx 1.6$\AA\ FWHM near the Ca{\sc ii} triplet at $\sim 8500$\AA\ (i.e., $R\approx 5300$). We bracketed the three individual exposures with arc lamp frames to ensure precise wavelength calibration.

\subsection{Radial velocity measurements}
We used the custom data reduction pipeline described in detail by \citet{ibata:11} to process our DEIMOS observations. Basic data processing steps are carried out first (debiasing, cosmic ray rejection, scattered light correction, flat-fielding, illumination and slit function correction, and fringe correction), followed by a wavelength calibration from the arc lamp frames that is finely corrected using the positions of the sky lines and is robust to better than $\approx 1-2$\,km\,s$^{-1}$ \citep[see][]{ibata:11}. A two-dimensional sky subtraction is performed, and the spectra are extracted without being resampled to give a set of observed pixel values, and associated uncertainties, for each target. Finally the pipeline measures radial velocities using a Bayesian methodology, whereby a simple template for the Ca{\sc ii} triplet is compared to the observations. A Markov Chain Monte Carlo algorithm is used to find the best-matching Doppler shift, and a robust uncertainty, for each target. The final radial velocity estimates are transformed to the heliocentric frame.

We were able to obtain measurements for $36$ targets. Individual velocity uncertainties are $\sim 8-12$\,km\,s$^{-1}$ for stars with $i_0 \ga 22.0$, decreasing to $\sim 4-5$\,km\,s$^{-1}$ for objects brighter than $i_0 \approx 21.5$. Because of the short integration time, most of our faint targets had insufficient signal-to-noise (S/N) for successful extraction of the spectra, or, in a few cases, for the velocity solution to adequately converge.

\begin{figure}
\centering
\includegraphics[height=76mm]{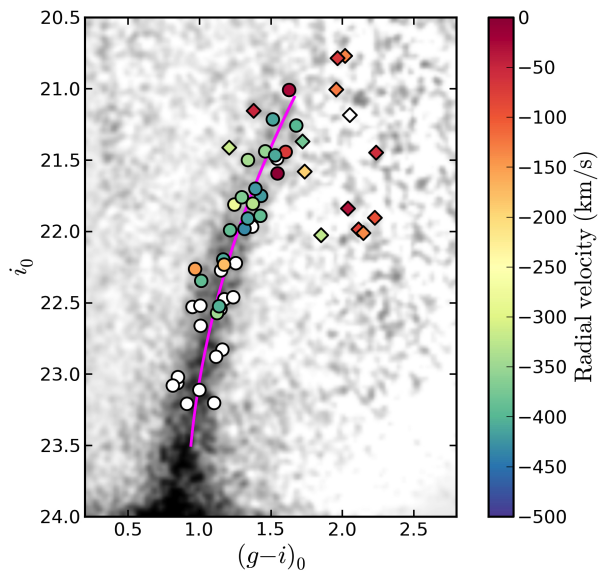}
\caption{Measured heliocentric velocities for our target stars as a function of position on the CMD. The underlying density map shows all stars inside the main body of the SWC after subtraction of the \citet{martin:13b} contamination model. Counts were conducted in $0.02 \times 0.02$ mag bins and then convolved with a Gaussian kernel of $\sigma = 0.03$ mag. The solid magenta line denotes the RGB parametrisation from Equation \ref{e:locus}. Overplotted points are our DEIMOS targets; circles are stars selected as SWC candidates, while diamonds are those selected as probable foreground. Individual velocities are indicated as per the colour scale; white denotes stars for which a measurement could not be obtained.}\label{f:cmd+vel}
\end{figure}

\section{Results \& Analysis}
\label{s:analysis}
\begin{figure}
\centering
\includegraphics[width=75mm]{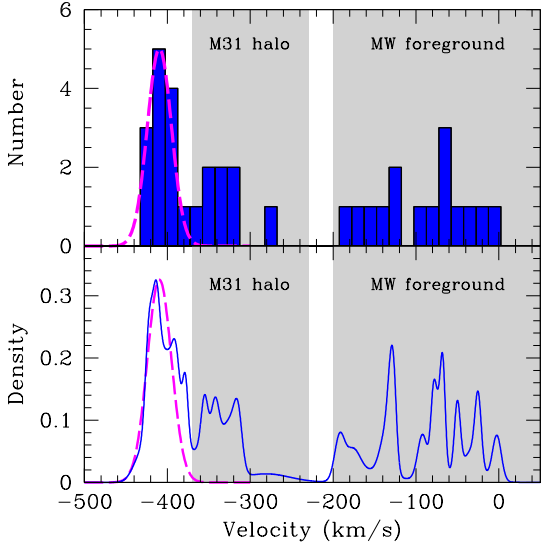}
\caption{{\bf Upper:} Histogram of heliocentric radial velocities for our targets. {\bf Lower:} The velocity distribution after applying an adaptive kernel estimator to remove binning effects, where the kernel is a Gaussian with $\sigma$ set to the velocity uncertainty for any given star. Both panels reveal a clear kinematic peak near $\sim -400$\,km\,s$^{-1}$ that we interpret as the SWC. The magenta dashed line denotes the most-likely model from our first analysis method --  the SWC profile is centred at $v_r = -409.5$\,km\,s$^{-1}$ and has a width given by the intrinsic dispersion $\sigma_v = 13.7$\,km\,s$^{-1}$ added in quadrature to the typical measurement uncertainty per star with $v \leq -370$\,km\,s$^{-1}$ of $\approx 6.5$\,km\,s$^{-1}$. The grey shaded areas indicate regions we expect to be occupied by the Milky Way foreground, and the M31 halo (i.e., $v_{\rm M31} \pm \sigma_{\rm M31}$ as described in Section \ref{s:analysis}).}\label{f:vhist}
\end{figure}

Figure \ref{f:cmd+vel} shows heliocentric radial velocities for our target stars as a function of position on the CMD. There is a clear kinematic signal 
for stars lying on or near the SWC RGB, quite distinct from those objects with positions more consistent with being members of the Milky Way foreground. This sub-grouping is also evident in Figure \ref{f:vhist}, which shows the velocity distribution. Stars for which $v \ga -200$\,km\,s$^{-1}$ (15 objects) belong to the Milky Way, while those with velocities $\la -250$\,km\,s$^{-1}$ (21 objects) are more likely to be part of M31.  Looking at the latter group, we see a relatively narrow peak just below $v \sim -400$\,km\,s$^{-1}$, which is well separated from the canonical M31 systemic motion $v_{\rm M31} = -301\pm 1$\,km\,s$^{-1}$ \citep{vandermarel:08}.  We interpret this peak as being due to the SWC.

As noted previously, M31 halo stars with similar metallicities to typical SWC members are not excluded by our photometric target selection. In general the two groups cannot easily be distinguished on the CMD, which means we must statistically separate them in velocity space. Because our sample is small, we elect to make some assumptions about the properties of the M31 halo to reduce the dimensionality of the problem. In terms of kinematics, we set the mean halo velocity to sit at the M31 systemic velocity. We also adopt a halo dispersion $\sigma_{\rm M31} \approx 70$\,km\,s$^{-1}$, obtained by extrapolating to $R_p = 90$ kpc the profile measured by \citet{chapman:06} inside $70$ kpc; note that the globular cluster system exhibits a comparable dispersion, $\sigma_{\rm M31} \approx 80$\,km\,s$^{-1}$, at this radius \citep{veljanoski:14}. 

Most critically, we estimate the degree of halo contamination in our sample as follows. First we recall that our original set of targets comprised two separate groups: SWC candidates selected from the RGB region of the CMD bounded by isochrones corresponding to $[$Fe$/$H$] \approx -1.8$ and $-1.1$, and likely Milky Way foreground objects chosen from outside this region. Eighteen of the $21$ stars in the final sample with $v < -250$\,km\,s$^{-1}$ come from the ``SWC candidate'' ensemble; the remaining three were obtained by chance in our set of ``foreground'' members. Considering the larger group first, \citet{bate:14} found that two-thirds of SWC stars have $-1.7 \la [$Fe$/$H$] \la -1.0$, along with an overall mean $V$-band surface brightness for the system of $\Sigma_V = 31.7\pm0.3$ mag arcsec$^{-2}$. \citet{ibata:14} found that at this radius, and in this quadrant, the M31 smooth halo component with $-1.7 \le [$Fe$/$H$] \le -1.1$ has $\Sigma_V \approx 32.9$ mag arcsec$^{-2}$.  Combining these two sets of measurements, and neglecting any second-order relationship between stellar metallicity and luminosity, suggests contamination at a level of roughly one M31 halo star per two SWC stars. That is, we expect $\approx 6$ of the $18$ ``SWC candidate'' stars with $v < -250$\,km\,s$^{-1}$ to be contaminants. Of the three serendipitous ``foreground'' stars with M31-like velocities, two lie well outside the populated range of the \citet{bate:14} MDF and are almost certainly not SWC members, while one is ambiguous. Overall, we therefore expect $\sim 8-9$ of the final sample of $21$ stars with $v < -250$\,km\,s$^{-1}$ to be M31 halo members, with the remainder belonging to the SWC.
 
We undertook two independent analyses to determine the kinematic properties of the SWC. For the first, we assume a Gaussian velocity profile centred on $v_r$ and with dispersion $\sigma_v$. The membership likelihood is given by
\begin{equation}
\mathcal{L} = \prod_j \left[ \frac{ 1 }{ \sqrt{2\pi(\sigma_v^2 + \sigma_j^2)} } \exp{\left( -\frac{ (v_r - v_j)^2}{2(\sigma_v^2 + \sigma_j^2)} \right)} \right]^{f( (g-i)_j , i_j )} 
\label{e:likelihood}
\end{equation} 
where $(v_j,\sigma_j)$ are the velocities and associated errors of the measured stars, and $f( (g-i)_j , i_j )$ is a weighting function\footnote{Note we drop the subscript `$0$' on magnitudes for clarity.} based upon the proximity of the stars to the RGB locus
\begin{equation}
(g-i) = 0.08229\ i^2 - 3.96354\ i + 48.63917\,.
\label{e:locus}
\end{equation}
The weighting function $f( (g-i)_j , i_j )$ is also taken to be a Gaussian, centred upon the locus and with a width of $\sigma_{\rm col}$; in the final analysis, two values of $\sigma_{\rm col} =$ 0.05 and 0.1 mag were considered, with no significant effect on the results.

\begin{figure}
\centering
\includegraphics[width=42mm]{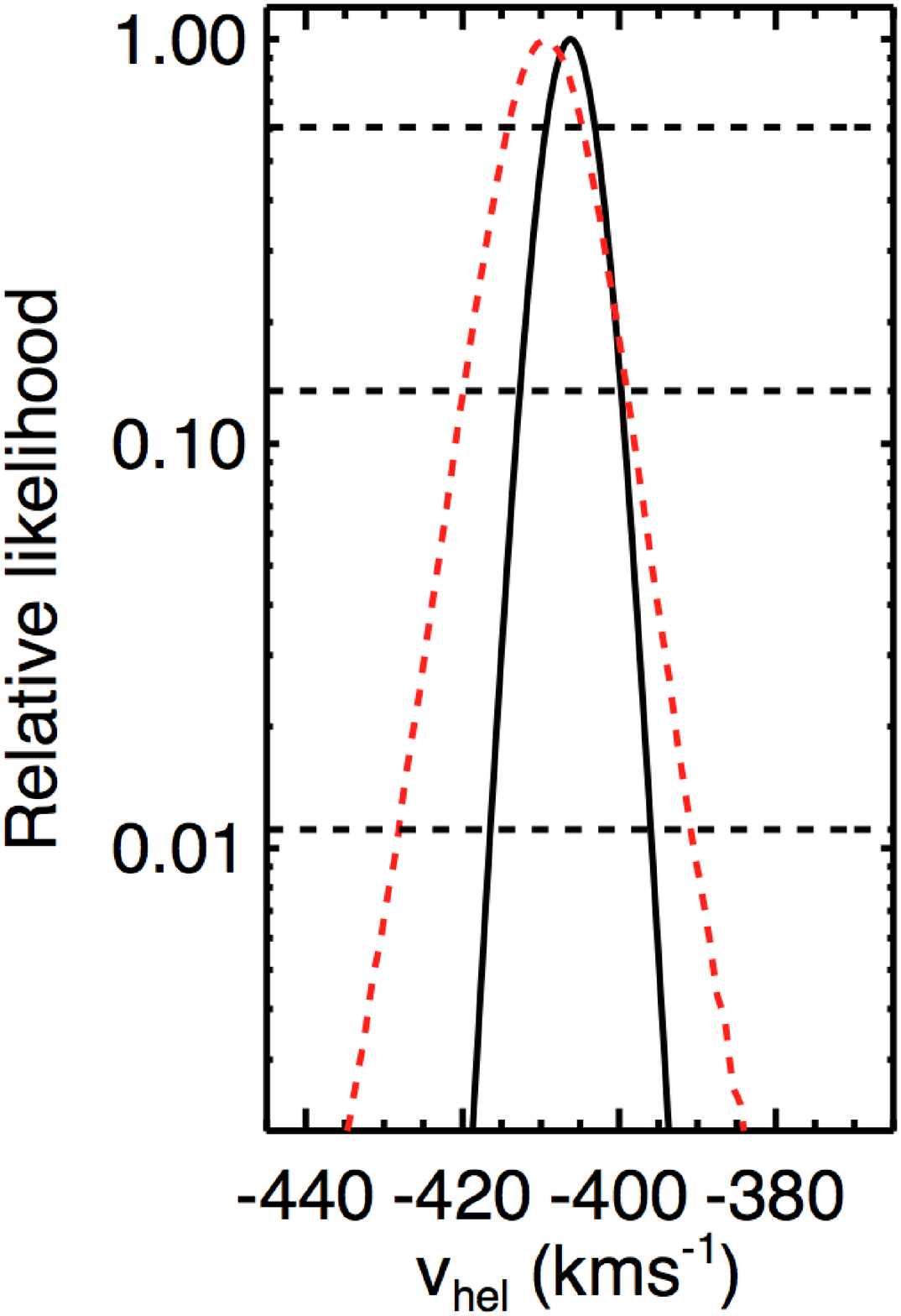}
\hspace{-1mm}
\includegraphics[width=38.1mm]{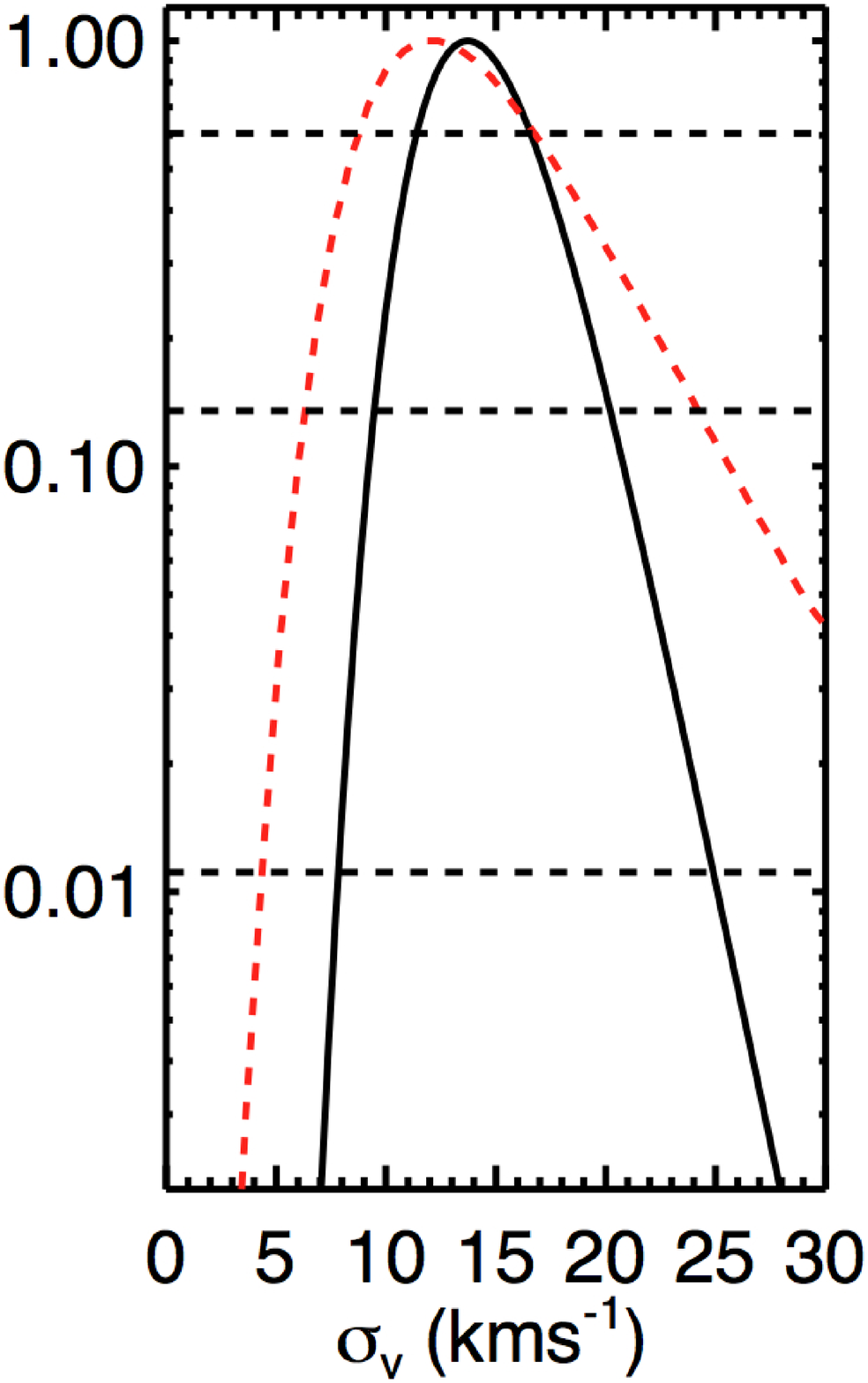}
\caption{Marginalised probability distribution functions for the SWC systemic velocity $v_r$ (left panel), and intrinsic dispersion $\sigma_v$ (right panel). Results from the first analysis method are plotted with a dashed red line, and those from the second method with a solid black line. The horizontal lines represent canonical 1, 2 and 3$\sigma$ confidence intervals, derived assuming Gaussian uncertainties.}\label{f:maxlike}
\end{figure}

We separated the SWC members from M31 halo contaminants by imposing a prior excluding stars with $v > -370$\,km\,s$^{-1}$. The velocity distribution to more negative values than the SWC peak near $\sim -400$\,km\,s$^{-1}$ falls off steeply, suggesting that the structure is relatively cold with $\sigma_v \la 20$\,km\,s$^{-1}$.  Thus our chosen prior sits $\ga 1\sigma_v$ from the SWC peak and $\sim 1\sigma_{\rm M31}$ from $v_{\rm M31}$, and results in 8 stars being assigned to the M31 halo -- in line with the expected degree of contamination outlined above. The search over parameter space yielded marginalised distributions, shown in Figure \ref{f:maxlike}, with $v_r = -409.5 \pm 5.8$\,km\,s$^{-1}$ and $\sigma_v = 13.7_{-4.2}^{+6.7}$\,km\,s$^{-1}$; these values correspond to the $50^{\rm th}$ percentiles, and the uncertainties to the 13.6 and 86.4 percentiles, respectively.  Note that the distribution in $\sigma_v$ is asymmetric; for completeness the peak value of the likelihood occurs at $(v_r,\,\sigma_v) = (-409.4,\,11.1)$\,km\,s$^{-1}$, whereas the peaks of the marginalised likelihoods occur at $v_r = -409.2$\,km\,s$^{-1}$ and $\sigma_v = 11.8$\,km\,s$^{-1}$ respectively.

Our second analysis method involved a more sophisticated treatment of the potential contaminants. For this we used a modified version of the algorithm developed by \citet{collins:13} to quantify the internal kinematics of M31 satellite dwarf galaxies. We refer the reader to that article for a full description; here it is sufficient to note that the method works by using a maximum likelihood approach to simultaneously fit the observed velocity distribution with several Gaussian profiles -- two for the Milky Way foreground, one for the M31 halo (with assumed properties negligibly different from those outlined above, $v_r = -308.8$\,km\,s$^{-1}$ and $\sigma_v = 96.3$\,km\,s$^{-1}$), and one for a substructure of arbitrary $(v_r , \sigma_v)$. The substructure profile matches Equation \ref{e:likelihood}, but with $f( (g-i)_j , i_j )$ replaced by $P_j$, representing the membership probability given a star's position both on the CMD and in velocity space\footnote{\citet{collins:13} also fold the distance from the dwarf galaxy centre into $P_j$; however, we ignore this as the DEIMOS field of view is much smaller than the spatial extent of the SWC.} -- i.e., $P_j \propto P_{\rm CMD} \times P_{\rm vel}$. The CMD probability is defined by placing a bounding box around the substructure RGB, outside of which $P_{\rm CMD}$ is zero, and within which it is determined from the smoothed density map. The velocity probability is obtained by iteratively considering the proximity of the star in velocity space to the contaminant profiles and the most likely location of the substructure profile, and taking into account the assumed fraction of stars in each population.

The marginalised probability distribution functions for $v_r$ and $\sigma_v$ from this approach are shown in Figure \ref{f:maxlike}. The algorithm clearly identifies the same peak as our first method, finding $v_r = -406.2 \pm 3.1$\,km\,s$^{-1}$ and $\sigma_v = 13.7_{-2.3}^{+2.9}$\,km\,s$^{-1}$, where the uncertainties reflect $1\sigma$ confidence intervals. Note the marginalized distributions are substantially narrower than those from the first method, due to the more careful modelling of the contaminating populations.

The systemic velocity we measure for our SWC field is relatively robust to changes in the assumed properties of the M31 halo.  For example, moving the velocity cut-off to $-340$\,km\,s$^{-1}$ in the first analysis method, or reducing the expected number of M31 halo contaminants to $4$ in the second method, leads to $v_r \approx -395 \pm 10$\,km\,s$^{-1}$. On the other hand, the inferred dispersion is quite sensitive to the estimated degree of contamination -- if this drops below $\sim 6$ stars (or the velocity cut-off is set above $\sim -350$\,km\,s$^{-1}$) then the dispersion blows out to $\ga 30$\,km\,s$^{-1}$. Related to this, Figure \ref{f:vhist} shows that the centre of the velocity distribution for likely M31 halo stars apparently sits closer to $\sim -340$\,km\,s$^{-1}$ than to the systemic velocity $v_{\rm M31} \approx -300$\,km\,s$^{-1}$. If our contamination assessment is accurate, this observation might hint at rotation in the M31 stellar halo. This would be in the same sense as that observed for the outer GC system by \citet{veljanoski:13,veljanoski:14}. While \citet{chapman:06} found little evidence for a rotating halo, we will revisit this tantalising possibility in a future contribution.

\section{Discussion}
\label{s:discuss}
Our Keck field is adjacent to the globular cluster PA-8, which has been shown to be cospatial in three dimensions with the SWC \citep[see][]{mackey:13,bate:14}. Two measurements of the velocity of PA-8 exist: $v_r = -411 \pm 4$\,km\,s$^{-1}$ \citep{mackey:13}, and $-416 \pm 8$\,km\,s$^{-1}$ \citep{veljanoski:14}. Within the uncertainties these are an excellent match to the systemic velocity of the central part of the SWC as derived here. Moreover, as noted previously the velocity of the cluster PA-14 \citep[$v_r = -363 \pm 9$\,km\,s$^{-1}$,][]{veljanoski:14}, which projects onto the southern outskirts of the SWC, is very close to that of the cold kinematic peak at $v_r = -373.5\pm 3.0$\,km\,s$^{-1}$ discovered serendipitously in a nearby field by \citet{gilbert:12}.

Taken together, this is compelling evidence that the globular clusters PA-8 and PA-14 are coherent in phase space with the stars that constitute the SWC, and hence that the GCs and the substructure share a common origin. Although this has previously been argued on a statistical basis \citep[see][]{mackey:10,mackey:13,veljanoski:14} our present results directly establish a physical link between the two. We thus confirm that the accretion of satellite dwarf galaxies deposits not only stars but also GCs into the halos of large galaxies, in a manner similar to that originally suggested for the Milky Way by \citet{searle:78} and demonstrated spectacularly with the discovery of the disrupting Sagittarius dwarf and its small cluster retinue.

\section*{Acknowledgments}
ADM is grateful for support by an Australian Research Fellowship (Discovery Project DP1093431) from the Australian Research Council (ARC).  GFL and NFB thank the ARC for support through Discovery Project DP110100678. GFL also gratefully acknowledges financial support through his ARC Future Fellowship (FT100100268).


\label{lastpage}
\end{document}